\documentclass[12pt,epsfig]{article}
\usepackage{graphicx}
\usepackage{amssymb}
\usepackage{mathrsfs}
\usepackage{amssymb,amsmath,bbm}
\usepackage{slashed}

\newcommand{\beq}[1] {\begin{equation}\label{#1} }
\newcommand{\eeq} {\end{equation} }
\newcommand{\bea}[1]{\begin{eqnarray}\label{#1} }
\newcommand{\eea}{\end{eqnarray}}

\begin{document}

\vspace*{-0.5cm}
\begin{flushright}
OSU-HEP-10-03
\end{flushright}
\vspace{0.5cm}

\begin{center}
{\Large {\bf Neutrino Masses from Fine Tuning} }

\vspace*{1.5cm} B.~N.~Grossmann\footnote{Email address:
benjamin.grossmann@okstate.edu}$^{,\dag}$, Z.~Murdock\footnote{Email address:
zeke.murdock@okstate.edu}$^{,\dag}$, and S.~Nandi\footnote{Email
address: s.nandi@okstate.edu}$^{,\dag}$

\vspace*{0.5cm}
$^{\dag}${\it Department of Physics and Oklahoma Center for High Energy Physics,\\
Oklahoma State University, Stillwater, Oklahoma, 74078\\}

\end{center}

\begin{abstract}

We present a new approach for generating tiny neutrino masses. The
Dirac neutrino mass matrix gets contributions from two new Higgs
doublets with their vevs at the electroweak (EW) scale. Neutrino
masses are tiny not because of tiny Yukawa couplings, or very heavy
($\sim 10^{14}\textrm{ GeV}$) right handed neutrinos. They are tiny
because of a cancelation in the Dirac neutrino mass matrix (fine
tuning). After fine tuning to make the Dirac neutrino mass matrix at
the $10^{-4}$ GeV scale, light neutrino masses are obtained in the
correct scale via the see-saw mechanism with the right handed
neutrino at the EW scale. The proposal links neutrino physics to
collider physics. The Higgs search strategy is completely altered.
For a wide range of Higgs masses, the Standard Model Higgs decays
dominantly to $\nu_L N_R$ mode giving rise to the final state
$\bar{\nu} \nu \bar{b} b$, or $\bar{\nu} \nu \tau^+\tau^-$. This can
be tested at the LHC, and possibly at the Tevatron.

\end{abstract}

\section{Introduction}

In the past decade, the existence of tiny neutrino masses of the
order of one hundredth to one tenth of an electron volt has been
firmly established through atmospheric, solar and reactor neutrino
experiments \cite{atmos}\cite{solar}\cite{reviews}. These masses are
a million or more times smaller than the corresponding charged
lepton masses. While the quark and charged lepton masses span many
orders of magnitude, the neutrino masses do not. The square roots of
the neutrino mass square differences, as obtained from the neutrino
oscillation experiments, lie within a factor of six of each other.
Also the quark mixing angles are very small, whereas two of the
neutrino mixing angles are large \cite{pdg}. These observations have
led to several unanswered questions. Why are the neutrino masses so
small compared to the corresponding charged lepton or quark masses?
Why is there such a large hierarchy among the charged fermion
masses, while there is practically no hierarchy among the neutrino
masses? Also, unlike the quark sector why are the mixing angles in
the neutrino sector large? Another related fundamental question is
whether neutrinos are Majorana or Dirac particles, and whether the
light neutrino spectrum exhibit a normal hierarchy or an inverted
hierarchy.

The most popular idea proposed so far for understanding the tiny
neutrino mass is the famous see-saw mechanism \cite{seesaw}. One
postulates the existence of very massive Standard Model (SM)
singlet right handed neutrinos with Majorana masses of the order of $M
\sim10^{14}\textrm{ GeV}$. The Yukawa couplings of the left-handed
(LH) neutrino to these heavy right-handed (RH) neutrinos then gives a
Dirac mass of the order of the charged lepton masses, $m_l$. As a
result, the left-handed neutrino obtains a tiny mass of the order of
$m^2_l/ M$. Although there are several indirect benefits for its
existence, there is no direct experimental evidence for such a heavy
particle. The mass scale is so high that no connection can be made
with the physics to be explored at the high energy colliders such as
the Tevatron and the LHC. It is important to explore other
possibilities to explain the tiny neutrino masses. Also, the see-saw
mechanism does not naturally lead to close values of the neutrino
masses as observed experimentally, though such close masses can be
arranged with the appropriate choice of the right handed Majorana
sector.

Recent astrophysical observation requires a tiny but non-zero value
of the cosmological constant, $\Lambda^{1/4} \simeq (10^{-4}$ eV).
This value is surprisingly close to the value of the light neutrino
masses required from the neutrino oscillation experiments, $\simeq
10^{-2}$--$10^{-1}$ eV. It has been exceedingly difficult to derive
such a tiny value of the cosmological constant, and there is some
acceptance that it may be fine tuned. The idea of Higgs mass also
being fine tuned has been explored leading to the so called ``Split
Supersymmetry'' \cite{ArkaniHamed:2004fb} with interesting
implications at the TeV scale that can be explored at the LHC.
Neutrino masses being in the same ballpark as the cosmological
constant, it is not unreasonable to assume that their values are
also fine tuned. The objective in this project is to adopt this
philosophy, build a concrete model realizing this scenario, and
explore its phenomenological implications, especially for the LHC.

In this work, we present a model in which the light neutrinos get
their masses from the usual see-saw mechanism, except the right
handed neutrino masses are at the EW scale. The neutrino Dirac
masses get contributions from two different Higgs doublets with
their vacuum expectation values (vevs) at the electroweak scale. The
neutrino masses are small not because of tiny Yukawa couplings or a
tiny vev of a new Higgs doublet \cite{Gabriel:2006ns}. In fact, we
take the Yukawa couplings to be of order one. The smallness of the
light neutrino masses are due to the cancellation in the Dirac
neutrino mass matrix, making $m_D\sim\mathcal{O}(10^{-4})\textrm{
GeV}$ and giving rise to light neutrino masses $m_{\nu} \sim m^2_D/
M$ where $M$ is the RH Majorana neutrino mass. Thus with $M$ in the
EW scale, we get the light neutrino masses in the correct range of
$10^{-2}$--$10^{-1}$ eV.

Our work is presented as follows: In section 2, we present the model
and the formalism. In section 3, we discuss the phenomenological
implications of the model, especially how it alters the usual SM
Higgs decay modes, and its implications for the Higgs search at the
LHC. Section 4 contains our conclusions.

\section{Model and Formalism}

\subsection{Our model}

Our model is based on the SM gauge symmetry $SU(3)_C\times SU(2)_L
\times U(1)_Y$ supplemented by a discrete $Z_2$ symmetry. In
addition to the SM fermions and the Higgs doublet $H$, we introduce
three RH neutrinos $N_{Ri}$, where $i=1,2,3$, and two additional
Higgs doublets $H_1$ and $H_2$,  with vevs at the EW scale. All the
SM particles are even under the $Z_2$ symmetry, while the three RH
neutrinos and the two new Higgs doublets $H_1$ and $H_2$ are odd
under $Z_2$. The $Z_2$ symmetry is softly broken by the bilinear
Higgs terms. With this symmetry, the Yukawa interactions are given
by
\begin{align}\label{yuk}
\mathcal{L}_{\text{SM Yukawa}}  & = \bar{q}_L y_u u_R \widetilde{H}
+ \bar{q}_L y_d d_R H + \bar{l}_L y_L e_R H + h.c.,
\end{align}
where the fermion fields represent three families, and $y_d$, $y_u$,
and $y_l$ represent three corresponding Yukawa coupling matrices.
\begin{align}\label{new}
\mathcal{L}_{\text{New Yukawa}}  &  = \bar{l}_L f_{1\nu} N_R
\widetilde{H}_1 +
\bar{l}_L f_{2\nu} N_R \widetilde{H}_2 + h.c.,\\
\mathcal{L}_{\text{Maj}} &= \frac{1}{2} M_{\text{Maj}} N^{T}_R C^{-1} N_R.
\end{align}
Note that from the above equations, the $6 \times 6$ neutrino mass
matrix is obtained to be
\begin{align}
M_{\nu}&=\begin{pmatrix}
0   &   m_{\text{D}}\\
(m_{\text{D}})^{\text{T}} & M_{\text{Maj}}
\end{pmatrix}.
\end{align}
The $3 \times 3$ Dirac  mass matrix is given by
\begin{align}\label{dirac}
m_{\text{D}}&= \frac{1}{\sqrt{2}}\left(f_{1\nu} v_1  +  f_{2\nu}
v_2\right).
\end{align}

Here $v_1$ and $v_2$ are the vevs of the new Higgs fields $H_1$ and
$H_2$. For the mass scales in which $m_{\text{D}}\ll M_{\text{Maj}}$, the
$3\times3$ light neutrino mass matrix is given by
\begin{align}\label{lightnumass}
m_{\nu}^{\textrm{light}}&=-m_{\text{D}}
M_{\text{Maj}}^{-1}(m_{\text{D}})^{\text{T}}.
\end{align}

Note that experimentally masses of the light neutrinos are in the
$10^{-1}$--$10^{-2}$ eV range. Thus with $M_{\text{Maj}}$ in the EW
scale, the matrix $m_{\text{D}}$ needs to be in the scale of
$10^{-4}$ GeV. Since the vevs $v_1$ and $v_2$ are in the EW scale,
we can get $m_{\text{D}}$ in the $10^{-4}$ GeV scale by assuming the
Yukawa couplings to be very tiny, of order $10^{-6}$. Such a path,
similar to the usual see-saw, will not lead to any interesting
implications for neutrino physics in the TeV scale. Instead we
assume that the Yukawa couplings $f_{1\nu}$ and $f_{2\nu}$ are
$\mathcal{O}(1)$, and these Yukawa couplings and vevs $v_1$ and
$v_2$ are fined tuned to get $m_{\text{D}}$ in the $10^{-4}$ GeV.
This is our approach to the smallness of the light neutrino mass
scale. As we will see, this gives interesting implication for
neutrino physics at the TeV scale, and can be explored at the LHC.

\subsection{Higgs potential}

Now we discuss the Higgs sector of the model. In addition to the
usual SM Higgs $H$ two other Higgs doublets $H_1$, $H_2$ are
required in this model. These two new Higgs doublets couple only to
the neutrinos, and this is imposed using the $Z_2$ symmetry. It is
the cancelation of contributions to the Dirac neutrino mass from
these two new doublets that enable the use of fine tuning.

We assume that the $Z_2$ symmetry is softly broken by the bilinear
terms in the Higgs Potential.
The two new doublets will mix with the SM Higgs doublet, and as we
will see, this will produce entirely new signals for the SM Higgs
boson decays. The Higgs potential is given by
\begin{align}
V_{\text{Higgs}}&=V^{(2)\text{even}}_{\text{Higgs}}+V^{(2)\text{odd}}_{\text{Higgs}}+V^{(4)\text{even}}_{\text{Higgs}},\\
V^{(2)\text{even}}_{\text{Higgs}}&=
\mu^2_{H} H^{\dag}H + \mu^2_{1} H^{\dag}_1 H_1 +\mu^2_{2} H^{\dag}_2 H_2+\mu^{2}_{12}(H^{\dag}_1 H_2+ h.c.),\\
V^{(2)\text{odd}}_{\text{Higgs}}&= \mu^2_{H1}(H^{\dag} H_1 + h.c.) +
\mu^2_{H2}(H^{\dag} H_2 + h.c.).
\end{align}
Note that the odd part of the potential breaks the $Z_2$ symmetry
softly.  This will have interesting implications for
the SM Higgs boson decays.
\begin{align}
\begin{split}
V^{(4)\text{even}}_{\text{Higgs}}&=
\lambda(H^{\dag} H)^2 + \lambda_1(H^{\dag}_1 H_1)^2+\lambda_2(H^{\dag}_2 H_2)^2\\
&\quad+\lambda_{1122}(H^{\dag}_1 H_1)(H^{\dag}_2 H_2)+\lambda_{HH12}(H^{\dag} H)\left(H^{\dag}_1 H_2+ h.c.\right)\\
&\quad+\lambda_{HH22}(H^{\dag} H)(H^{\dag}_2 H_2)+\lambda_{1112}(H^{\dag}_1 H_1)\left(H^{\dag}_1 H_2+ h.c.\right)\\
&\quad+\lambda_{HH11}(H^{\dag} H)(H^{\dag}_1 H_1)+\lambda_{2212}(H^{\dag}_2 H_2)\left(H^{\dag}_1 H_2+ h.c.\right)\\
&\quad+\lambda_{12}(H^{\dag}_1 H_2+h.c.)^2+\lambda_{H1}(H^{\dag} H_1+h.c.)^2 + \lambda_{H2}(H^{\dag} H_2+h.c.)^2\\
&\quad+\lambda_{H1H2}\left(H^{\dag} H_1+ h.c.\right)\left(H^{\dag} H_2+h.c.\right).\\
\end{split}
\end{align}

Since there are three Higgs doublets, after EW symmetry breaking,
there will remain a pair of charged Higgs ($H^{\pm},H^{\prime\pm}$),
five neutral scalar Higgses ($h^\prime,h^{\prime}_1, h^{\prime}_2,
H^{\prime}_1, H^{\prime}_2$), and two neutral pseudoscalar Higgses
($A^{\prime}_1, A^{\prime}_2$).  Due to the breaking of the $Z_2$
symmetry, there is mixing within each of these three groups of
Higgses (but not between groups). We denote the mass eigenstates of
the five neutral Higgses by $h$, $h_{10}$, $h_{20}$, $H_{10}$, and
$H_{20}$.

\subsection{Mixing between the light and heavy neutrinos}

In our model, we are considering a scenario in which the three RH
neutrinos have masses in the EW scale with $\mathcal{O}(1)$ Yukawa
couplings with the light LH neutrinos. They will also mix with the
light neutrinos, and thus will participate in the gauge
interactions. LEP has searched for such RH neutrinos. Before we
discuss these constraints, let us first consider the mixing between
the light neutrinos and the RH neutrinos. Using the observed values
of the light neutrino masses and mixings, we can make a reasonable
estimate of the mixing between the LH and RH neutrinos as follows.
We use the normal hierarchy for the light neutrino masses with the
values
\begin{align}
m^{\text{light}}_{\nu\text{Eigenvalues}}&=\text{Diag}(m_{\nu_1},m_{\nu_2},m_{\nu_3})=\text{Diag}(0,8.71,49.3)\times10^{-12}\textrm{
GeV}.
\end{align}
The mixing matrix $R_{\nu\nu}$ follows the standard parametrization.
The angles $\theta_{12},\theta_{23}$ are the central values, and
$\theta_{13}$ is the maximal value allowed by current experiment
\cite{chooze}.
\begin{align}
(\theta_{12},\theta_{23},\theta_{13})&=(0.601,0.642,0.226),
\end{align}
\begin{align}
R_{\nu\nu}&=
\begin{pmatrix}
0.804   &   0.551   &   0.223\\
-0.563  &   0.585   &   0.584\\
0.190   &   -0.595  &   0.781
\end{pmatrix}.
\end{align}
The three possible CP-violating phases are assumed to be zero. From
the above mass eigenvalues and the mixing matrix, we can calculate
the light neutrino mass matrix using
\begin{align}
(R_{\nu\nu})^{\textrm{T}}m_{\nu}^{\textrm{light}}R_{\nu\nu}=m_{\nu\textrm{Eigenvalues}}^{\textrm{light}}.
\end{align}
For simplicity, we assume that  the $3 \times 3$ RH Majorana mass
matrix $M_{\text{Maj}}$ to be proportional to the unit matrix,
\begin{align}
M_{\text{Maj}}&=\text{Diag}(M,M,M).
\end{align}
As a consequence of this choice for $M_{\text{Maj}}$ and having a
symmetric $m_{\text{D}}$, the mixing matrix among only the
generations of heavy neutrinos is equivalent to the mixing matrix
among only the generations of light neutrinos $R_{NN}=R_{\nu\nu}$.
Using the above numbers and choosing $M=100\text{ GeV}$, we can now
calculate numerically the $3\times 3$ Dirac neutrino mass matrix
from the equation
\begin{align}
m_\nu^{\text{light}}&=-m_{\text{D}}M^{-1}_{\text{Maj}}m_{\text{D}}^\text{T}.
\end{align}
There are four sets of real solutions for $m_{\text{D}}$. Only two
sets of solutions are shown in Eqs. (\ref{mDset1}) and
(\ref{mDset2}). The other two are just the negatives of these two
sets.
\begin{align}
\label{mDset1}
m_{\text{D}}^{\text{Set 1}}&=
\begin{pmatrix}
-1.25   &   -1.87   &   -0.267  \\
-1.87   &   -3.42   &   -2.20   \\
-0.267  &   -2.20   &   -5.36
\end{pmatrix}\times10^{-5}\text{ GeV},\\
\label{mDset2}
m_{\text{D}}^{\text{Set 1}}&=
\begin{pmatrix}
-0.543  &   -0.0280 &   2.20\\
-0.0280 &   1.40    &   4.25\\
2.20    &   4.25    &   3.27
\end{pmatrix}\times10^{-5}\text{ GeV}.
\end{align}

Using the solutions for $m_{\text{D}}$ and $M_{\text{Maj}}$, we can
now use the full $6 \times 6$ neutrino mass matrix and calculate the
full mixing matrix $Q$ and the mixing angles between the heavy and
light neutrinos.
\begin{align}
M^{\text{Full}}&=\begin{pmatrix}
0_{3\times 3}&m_{\text{D}}\\
m_{\text{D}}^{\text{T}}&M_{\text{Maj}}
\end{pmatrix},&
Q^{-1}M^{\text{Full}}Q=M^{\text{Full}}_{\text{Eigenvalues}}.
\end{align}
It turns out that
\begin{align}
Q&\approx
\begin{pmatrix}
R_{\nu\nu}&Q_{\nu N}\\Q_{N\nu}&R_{NN}
\end{pmatrix},&
Q_{\nu N}&\approx Q_{N\nu}.
\end{align}
For solution set 1, the full rotation matrix is
\begin{align}
Q^{\textrm{Set 1}}&=\begin{pmatrix}
\begin{matrix}
0.80   &0.55  &0.22\\
-0.56  &0.59  &0.58\\
0.19   &-0.60 &0.78
\end{matrix}\phantom{\times10^{-7}}&\begin{pmatrix}
3.1\times10^{-3}  &1.6   &1.6\\
3.513\times10^{-3}  &1.7   &4.1\\
-2.5\times10^{-3} &-1.8  &5.5
\end{pmatrix}\times10^{-7}\\
\begin{pmatrix}
1.1\times10^{-4}   &-1.6  &-1.6\\
1.3\times10^{-4}   &-1.7  &-4.1\\
-8.8\times10^{-5}  &1.8   &-5.5
\end{pmatrix}\times10^{-7}&\begin{matrix}
0.81  &0.55  &0.22\\
-0.56 &0.59  &0.58\\
0.19  &-0.60 &0.78
\end{matrix}\phantom{\times10^{-7}}
\end{pmatrix}.
\end{align}
For solution set 2, the full rotation matrix is
\begin{align}
Q^{\textrm{Set 2}}&=\begin{pmatrix}
\begin{matrix}
0.80   &0.55  &0.22\\
-0.56  &0.59  &0.58\\
0.19   &-0.60 &0.78
\end{matrix}\phantom{\times10^{-7}}&\begin{pmatrix}
4.1\times10^{-3}   &1.6   &-1.6\\
3.9\times10^{-3}   &1.7   &-4.1\\
-5.7\times10^{-3}  &-1.8  &-5.5
\end{pmatrix}\times10^{-7}\\
\begin{pmatrix}
-8.9\times10^{-4}  &-1.6  &1.6\\
-7.9\times10^{-4}  &-1.7   &4.1\\
1.4\times10^{-3}   &1.8   &5.5
\end{pmatrix}\times10^{-7}&\begin{matrix}
0.80   &0.55  &0.22\\
-0.56  &0.590  &0.58\\
0.19   &-0.60 &0.78
\end{matrix}\phantom{\times10^{-7}}
\end{pmatrix}.
\end{align}
\renewcommand{\arraystretch}{1.2}
\begin{table}[tb]
\begin{center}
\caption{Mixing angles between the light neutrinos (subscripts
$1,2,3$) and the heavy neutrinos (subscripts $4,5,6$).}
\label{tbl:nuangles}
\begin{tabular}{c|ccccccccc}\hline\hline
$\times10^{-7}$&$\theta_{14}$&$\theta_{15}$&$\theta_{16}$&$\theta_{24}$&$\theta_{25}$&$\theta_{26}$&$\theta_{34}$&$\theta_{35}$&$\theta_{36}$\\\hline
Set 1   &$1.2$ &$1.9$ &$0.26$ &$1.9$ &$3.4$  &$2.2$  &$0.26$  &$2.2$  &$5.3$\\
Set 2   &$0.55$ &$0.36$ &$-2.2$ &$0.034$ &$-1.4$ &$-4.2$ &$-2.2$ &$-4.2$ &$-3.2$\\
\hline\hline
\end{tabular}
\end{center}
\end{table}
As can be seen on Table~\ref{tbl:nuangles}, the mixing between the
heavy and light neutrinos is extremely small.

\section{Phenomenological implications}

In this section, we discuss the phenomenological implications of our
model. We are considering RH neutrinos at the EW scale. Their mass can
be below the $W$ boson mass. Thus they can be searched for at LEP,
Tevatron, and at the LHC. First we discuss the
constraints that already exist from the search at LEP.

\subsection{LEP constraints}

Searches for $N_R$ have been conducted at LEP in the channel $e^+
e^-  \rightarrow  Z \rightarrow N_R \nu_l $, with $N_R$ subsequently
decaying to $W^+ e^- $ or $Z \nu $. This experiment puts limits on
the mixing angle $\theta$ between the heavy and the light neutrinos
$\sin^2 \theta < 10^{-4}$ for $3$ GeV $< M_N < 80$ GeV, and  $\sin^2
\theta < 0.1$ for $M_N > 80$ GeV \cite{Achard}. As we discussed in
previous section, the mixing angles $\theta$ between the light and
heavy neutrinos are extremely small, ranging between $\sim 10^{-8}$
to $10^{-6}$. Thus, in our model, LEP constraints allow small masses
for the heavy Majorana neutrinos.

\subsection{Higgs decays and Higgs signals}

In our model, the Yukawa couplings between the light neutrinos, the
heavy Majorana neutrinos and the new Higgs fields $H_1$ and $H_2$
are $\mathcal{O}(1)$. The SM Higgs $H$ mixes with the new Higgses,
and these mixings are naturally large. Thus, for $M_N < M_h$, the SM
Higgs will dominantly decay to a light $\nu$ and $N_R$, as soon as
this decay mode becomes kinematically allowed, because the coupling
for this decay mode is much larger than the usually dominant
$\bar{b} b $ mode, or even the $WW$ mode.  The branching ratios for
the various Higgs decay modes are shown in Fig. 1 for $M_N = 80$ GeV,
for the Yukawa couplings, and $f_{1\nu} = f_{2\nu} =1$. As can be
seen from the plot, as soon as the decay mode $h \rightarrow \nu
N_R$ becomes kinematically allowed, this mode totally dominates over
the usual $\bar{b} b $ mode, and is larger than the usually dominant
$WW$ mode even beyond the $WW$ threshold. Thus in our model, the SM
Higgs decay mode is greatly altered.  A second plot is shown for
Yukawa couplings equal to $1/14$ in Fig. \ref{h-2x_0.25}
\begin{figure}[t]
    \begin{center}
        \includegraphics{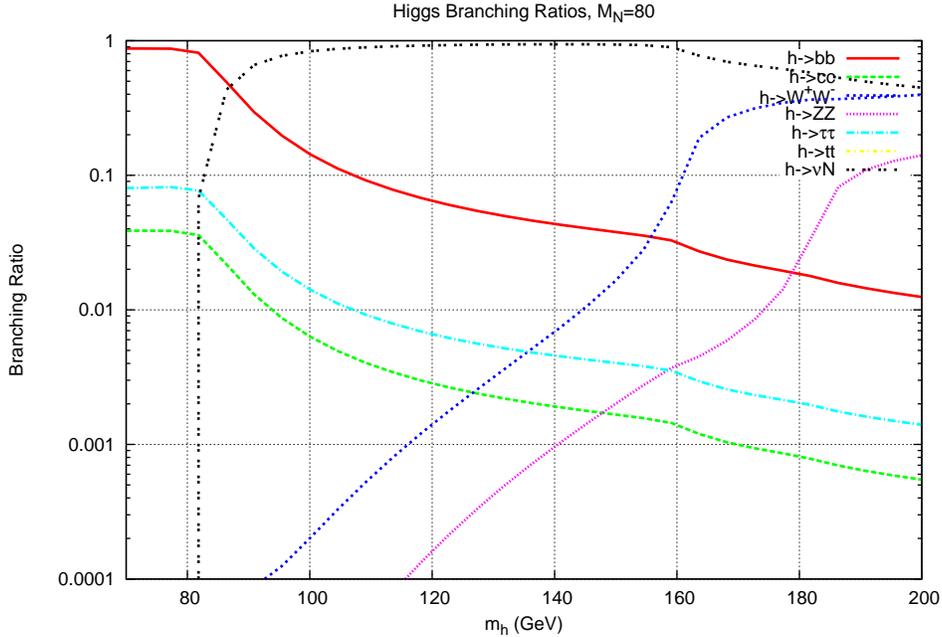}
        \caption{Branching ratios of $h\rightarrow2x$ with the coupling between $N_R$ and $\nu_L$, $f_{1\nu}+f_{2\nu}=1$.}
        \label{h-2x}
    \end{center}
\end{figure}
\begin{figure}[t]
    \begin{center}
        \includegraphics{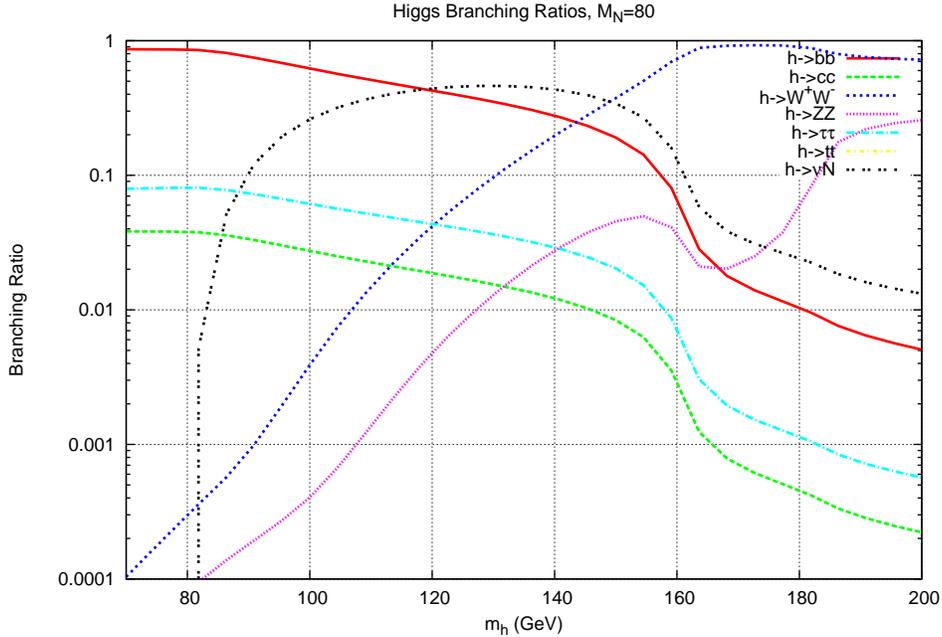}
        \caption{Branching ratios of $h\rightarrow2x$ with $f_{1\nu}+f_{2\nu}=\frac{1}{7}$.}
        \label{h-2x_0.25}
    \end{center}
\end{figure}

At hadron colliders, the SM Higgs boson is dominantly produced via
gluon fusion with the top quark in the loop. In our model, because
of the mixing of $H$ with $H_1$ and $H_2$, the lightest mass neutral
scalar Higgs decays dominantly as $h \rightarrow \nu N_R$.
\begin{figure}[t]
    \begin{center}
        \includegraphics[scale=0.7]{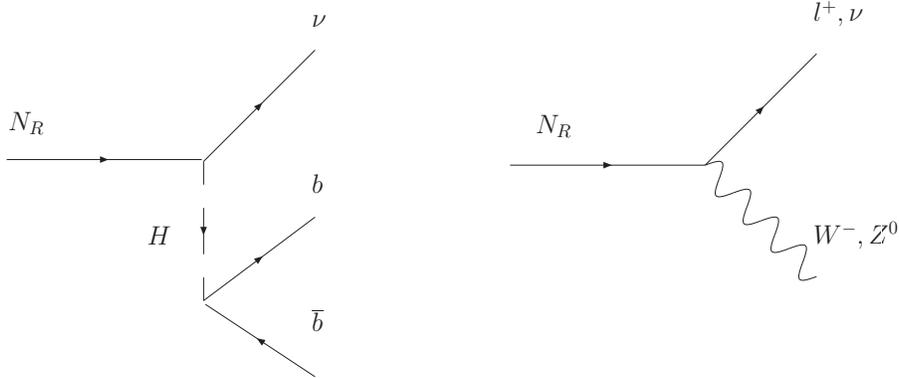}
        \caption{Decay modes of $N_R$.}
        \label{feyndiag}
    \end{center}
\end{figure}
The final state signal will depend on the decay modes of $N_R$. Two
of the allowed decay modes of $N_R$ are shown in Fig.
\ref{feyndiag}. The 3-body decay mode $N_R \rightarrow \nu \bar{b}
b$ is completely dominant over the 2-body decay mode $lW$ or $\nu
Z$. This is because the 2-body decay is suppressed by the tiny
mixing angle, $\theta \sim 10^{-6}$ or smaller.  Thus the final
state signals for the Higgs bosons at the LHC, in our model, is
$\bar{\nu} \nu \bar{b} b$. Collider signals will include large
missing energy and 2 hard b-jets.

Using Madgraph, we generated events for $p p \rightarrow \bar{\nu}
\nu \bar{b} b$ in the SM for LHC at 14 TeV, 7 TeV, and Tevatron.
Using the cuts $\slashed{E}_T > 30$ GeV, and the $p_T$ for each
b-jet to be greater than $20$ GeV, we find the cross section to be
$\sim 13$ pb, for the LHC at 14 TeV.  This provides a reasonable
estimate of the background.  The cross section for Higgs production
at the LHC at $14$ TeV is $\sim 50$ pb for a 120 GeV Higgs. For a
large mass range of the Higgs boson in our model, the branching
ratio, $BR(h \rightarrow \nu N_R) \sim 100\%$. Thus, prior to any cuts on the signal, this mode
is observable at the LHC, and stands out over
the SM background.  A summary for different energies is given in
Table \ref{collidersearch}. The Higgs production at the Tevatron is
taken from \cite{Baglio}.  For the LHC we used \cite{Suarez}.

\begin{table}[ht]
\begin{center}
\caption{Collider Searches for $m_h=120\text{ GeV}$.}
\label{collidersearch}
\begin{tabular}{c c c c}
\hline\hline
Collider & $\sqrt{s}$ Energy & Background & Signal \\ 
\hline
LHC      & 14 TeV        & 13 pb      & 50 pb \\
LHC      & 7 TeV         & 2.4  pb    & 30  pb \\
Tevatron & 2 TeV         & 240  fb      & 1  pb \\
\hline\hline
\end{tabular}
\end{center}
\end{table}


\subsection{$ZH\rightarrow \nu \bar{\nu} b \bar{b}$ Search at Tevatron}

The D0 collaboration at the Tevatron has searched for the SM Higgs boson in the $ZH\rightarrow \nu \bar{\nu} b \bar{b}$ channel using 5.2 $\mathrm{fb}^{-1}$ of data.  With both $b$'s being tagged and a $\slashed{E}_T >40 \textrm{ GeV}$ and $p_T > 20 \textrm{ GeV}$ for the b-jets, they expect about 5 events for the $ZH$ mode.  However, there is a large SM background arising mainly from $W$ + jets, $Z$ + jets, and $t \bar{t}$.  The estimated background with these cuts is about $538 \pm 93$ events, while they observe 514 events.  Thus the SM signal from $ZH$ production for this $\nu \bar{\nu} b \bar{b}$ mode is not observable with the current Tevatron data.  However in our model, depending on the $BR(h\rightarrow \nu N_R)$, this signal is much larger and may be observable, especially as luminosity accumulates in the coming year.  Other possibilities for our model are that the RH neutrinos could decay via a charged Higgs.
\subsection{$N_R$ Decays via Charged Higgs}
For a sufficiently light $m_{H^\pm}<250\textrm{ GeV}$, the decay
$N_R\rightarrow \nu_\tau \tau^+ \tau^-$ via a charged Higgs becomes
important.  Taking the Yukawa couplings to be order one, and the
mixing to be maximal between the three Higgs doublets, the decay
rates for the $N_R$ decays are shown in Table \ref{brs}.  Taking the
tau $p_T > 20 \textrm{ GeV}$ and $\slashed{E}_T >30 \textrm{ GeV}$,
the cross section for $p\bar{p}\rightarrow \nu \bar{\nu} \tau^+
\tau^-$ at the Tevatron is 45 fb (123 fb at the LHC for 7 TeV
collisions).  This background is much smaller than $pp\rightarrow
\bar{b} b\nu \bar{\nu}$ background, as it is a leptonic (not QCD)
process. This signature, two high $p_T$ taus plus missing energy,
may be easier to see.
\begin{figure}[t]
    \begin{center}
        \includegraphics[scale=0.7]{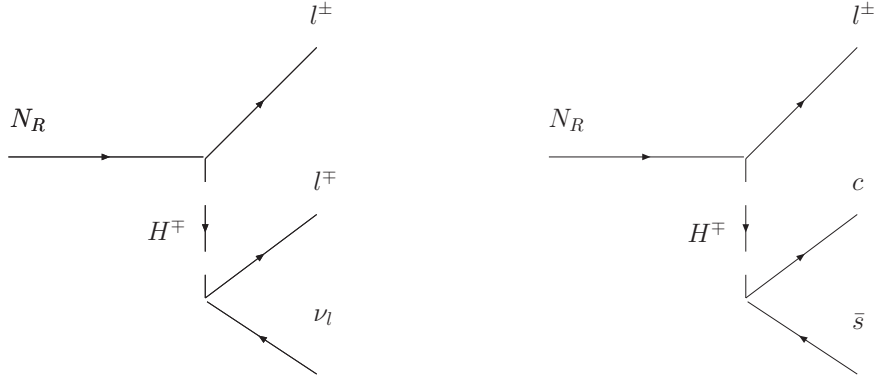}
        \caption{Decay modes of $N_R$ through a charged Higgs $H^\mp$.}
        \label{feyndiag2}
    \end{center}
\end{figure}

\begin{table}[ht]
\begin{center}
\caption{Decay Rates of $N_R$ for $M_N = 80\textrm{ GeV}$, $M_h = 120\textrm{ GeV}$.}
\label{brs}
\begin{tabular}{l l c l}
\hline\hline
Decay Mode                  & $\Gamma(N_R\rightarrow 3x)$ (GeV)     & $m_{H^\pm}$ (GeV) & BR \\ 
\hline
$N_R\rightarrow \nu b \bar{b}$          & $1.56\times 10^{-9}$            & 200           & 43.8\% \\
$N_R\rightarrow \nu_\tau \tau^+ \tau^-$     & $1.32\times 10^{-9}$                    & 200           & 37.0\% \\
$N_R\rightarrow \tau c \bar{s}\ (\textrm{or}\ \bar{c} s)$   & $5.80\times 10^{-10}$   & 200           & 16.3\%\\
$N_R\rightarrow \nu c \bar{c}$          & $6.60\times 10^{-11}$           & 200           & 1.85\%\\
$N_R\rightarrow \nu_\mu \mu^+ \mu^-$        & $4.00\times 10^{-11}$           & 200           & 1.12\%\\
\hline
$N_R\rightarrow \nu b \bar{b}$          & $1.56\times 10^{-9}$            & 250           & 63.6\% \\
$N_R\rightarrow \nu_\tau \tau^+ \tau^-$     & $5.62\times 10^{-10}$               & 250           & 22.9\% \\
$N_R\rightarrow \tau^- c \bar{s}\ (\textrm{or}\ \tau^+ \bar{c} s)$  & $2.26\times 10^{-10}$   & 250           & 9.21\%\\
$N_R\rightarrow \nu c \bar{c}$          & $6.60\times 10^{-11}$           & 250           & 2.69\%\\
$N_R\rightarrow \nu_\mu \mu^+ \mu^-$        & $2.45\times 10^{-11}$           & 250           & 1.65\%\\
\hline\hline
\end{tabular}
\end{center}
\end{table}

\section{Conclusions}

We have proposed a new approach for understanding of the tininess of
the light neutrino masses. We extend the SM gauge symmetry by a
discrete $Z_2$ symmetry, and the particle content by adding three
right handed neutrinos and two additional Higgs doublets. These new
Higgs doublets couple only to the neutrinos. The tiny neutrino
masses are generated via the see-saw mechanism with the right handed
neutrino mass matrix at the EW scale, and the Dirac neutrino mass
matrix at the $10^{-4}$ GeV scale. The Dirac neutrino mass matrix
gets contribution from the two new EW Higgs doublets with vevs at
the EW scale. The Yukawa couplings are of order one, and the two EW
contributions are fine tuned to achieve the Dirac neutrino mass
matrix at the $10^{-4}$ GeV level. The model links neutrino physics
to collider physics at the TeV scale. The SM Higgs decays are
drastically altered. For a wide range of the Higgs mass, it decays
dominantly to $\nu_L N_R$ giving rise to the final state $\bar{\nu}
\nu \bar{b} b$, or $\bar{\nu} \nu \tau^+\tau^-$. This can be tested
a at the LHC and possibly at the Tevatron.

\section*{Acknowledgments}

We thank K. S. Babu, W.A.  Bardeen, B. Dobrescu, C. T. Hill and A.
Khanov for useful discussions. Part of this work was done during our
visit to Fermilab in Fall, 2009. We thank the Fermilab Theory Group
for warm hospitality and support during this visit. This work is
supported in part by the United States Department of Energy, Grant
Numbers DE-FG02-04ER41306 and DE-FG02-04ER46140.

\end{document}